\documentclass[journal]{IEEEtran}

\ifCLASSINFOpdf
\else
   \usepackage[dvips]{graphicx}
\fi
\usepackage{url}
\usepackage{graphicx}

\usepackage{booktabs}
\usepackage{multirow}
\usepackage{makecell}
\usepackage{tabularx}
\usepackage{pifont}
\usepackage{bbding}
\usepackage{cite}

\begin{document}

% \title{An overview of current audio codec-based language models}
\title{Towards audio language modeling - an overview}

% \author{{Xuanjun Chen}$^{*}$, {Haibin Wu}$^{*}$, {Yi-Cheng Lin}$^{*}$, {Kai-wei Chang}$^{*}$, {Hung-yi Lee}
\author{Haibin Wu$^{1}$, Xuanjun Chen$^{1*}$, Yi-Cheng Lin$^{1*}$, Kai-wei Chang$^{1}$, Ho-Lam Chung$^{1}$, \\
Alexander H. Liu$^{2}$, Hung-yi Lee$^{1}$ \\
% $^{1}$National Taiwan University \\
% $^{2}$Massachusetts Institute of Technology \\
\thanks{$^{*}$Equal second contribution. $^{1}$ National Taiwan University. $^{2}$Massachusetts Institute of Technology. }
}

\maketitle

%%%% Xuanjun version @ 2024.02.19
\begin{abstract}
Neural audio codecs are initially introduced to compress audio data into compact codes to reduce transmission latency.
Researchers recently discovered the potential of codecs as suitable tokenizers for converting continuous audio into discrete codes, which can be employed to develop audio language models (LMs).
Numerous high-performance neural audio codecs and codec-based LMs have been developed.
The paper aims to provide a thorough and systematic overview of the neural audio codec models and codec-based LMs.
\end{abstract}

\begin{IEEEkeywords}
Neural codec, codec-based language model
\end{IEEEkeywords}

\IEEEpeerreviewmaketitle

\begin{figure*}[ht]
    \centering
    \includegraphics[width=1.8\columnwidth]{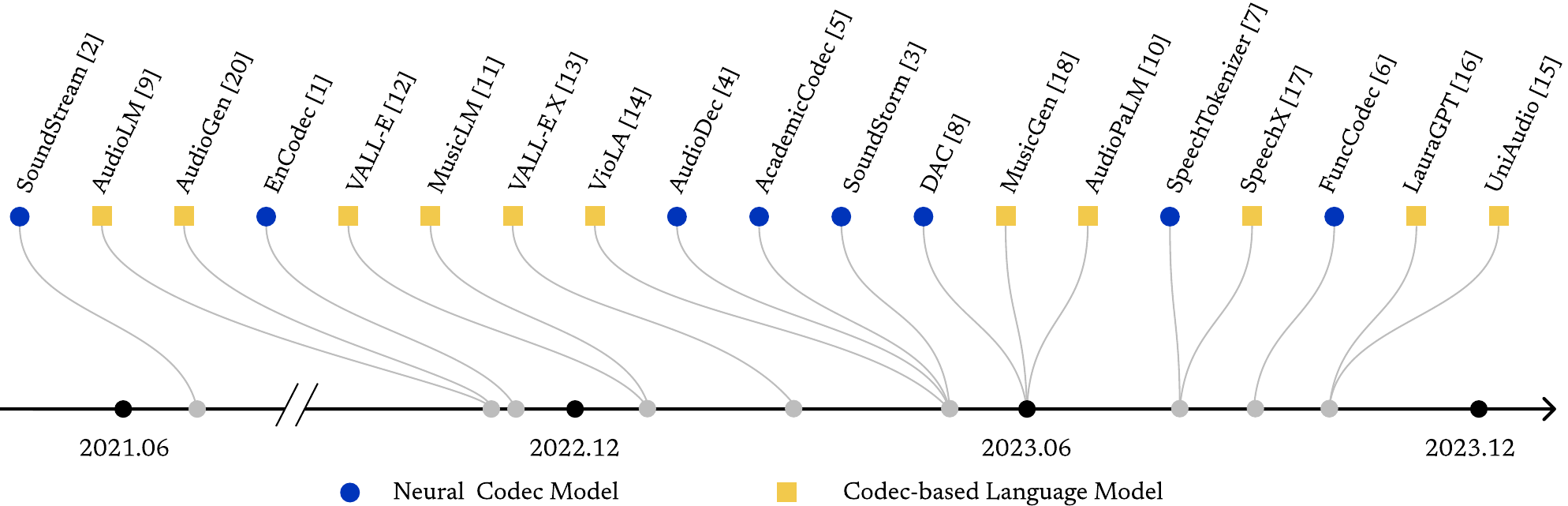}
    \caption{Timeline of current neural codec models and codec-based language models.}
    \label{fig:framework}
    \vspace{-10pt}
\end{figure*}
\section{Introduction}
\label{sec:intro}

Neural audio codec models were first introduced to compress audio for efficient data transmission. 
The encoder converts the audio into codec codes, which are then transmitted. 
The receiver then uses the codec decoder to reconstruct the audio using the received codes.

Language modeling has proven to be highly successful in the field of Natural Language Processing (NLP). 
Audio data encompasses not only textual content but also rich information about speaker timbre, emotion, and general audio, offering deeper possibilities for language model applications.
Researchers, especially those in large companies with significant computational resources, recently leverage the potential of neural codecs 
\cite{defossez2022high, zeghidour2021soundstream, borsos2023soundstorm, wu2023audiodec, yang2023hifi, du2023funcodec, zhang2023speechtokenizer, kumar2023high} 
as suitable tokenizers for converting continuous audio into discrete codes, which can be employed to develop audio language models (LMs) \cite{borsos2023audiolm,rubenstein2023audiopalm,agostinelli2023musiclm,wang2023neural,zhang2023speak,wang2023viola,yang2023uniaudio,chen2023lauragpt,wang2023speechx,copet2023simple,lan2023stack,kreuk2022audiogen}.
The current codec-based language models and codec models are summarized in Figure~\ref{fig:framework}.
These findings promptly garnered the community's attention, sparking a fervor for developing codecs tailored to audio language modeling. 
Numerous high-performance neural audio codec models and audio LMs have been developed.

An ideal codec should maintain content while preserving paralinguistic and speaker-related information. 
Similarly, a universal audio language model should be able to generalize across various audio types, such as speech, music, and general audio, covering a wide range of applications. 
The arms race in developing codecs and audio LMs is still ongoing.

Given the significant advancements in codecs and audio language models over the past three years as shown in Figure~\ref{fig:framework}, there has yet to be a comprehensive review comparing them and providing inspiration to the community.
In this study, we aim to fill this research gap by thoroughly reviewing and comparing various existing neural codec models and audio codec-based language models.
Firstly, we specifically conduct an in-depth analysis of six representative open-source neural codec models to cover their training methodologies, implementation settings, and training data.
Secondly, we expand our analysis to include eleven diverse codec-based language models, examining how they utilize the codecs and the tasks to which they can be applied. 
Through this comprehensive review, we aim to offer the community insights into the diverse methodologies and potential directions in the field of neural codecs and codec-based language modeling.

\section{Comprehensive comparison for neural audio codec models}
Codec models aim to compress and decompress speech signals efficiently. 
Traditional codecs are developed based on psycho-acoustics and speech synthesis \cite{valin2012rfc, dietz2015overview}. 
Recently, the neural codec models demonstrated highly effective for compression and signal reconstruction, outperforming traditional codecs.
Considering the broad spectrum of codec models within the research community, each trained with its distinct configurations and training techniques, there is a clear need for a thorough examination that covers the training methodologies, implementation settings, and training data employed across these codec models.
The six codec models have distinct training details, resulting in a collection of fifteen different codec models, as summarized in Table~\ref{tab:codec_info}.

\subsection{Brief method overview for codecs}
\label{subsec:comprehensive_review}

SoundStream \cite{zeghidour2021soundstream} stands as one of the pioneering implementations of neural codec models, embodying a classic neural codec architecture comprising encoder, quantizer, and decoder modules.
It utilizes the streaming SEANets \cite{tagliasacchi2020seanet} as its encoder and decoder. 
The quantizer incorporates a speech enhancement system with a Residual Vector Quantization (RVQ) \cite{kumar2019melgan, zeghidour2021soundstream} bottleneck to obtain parallel token streams. 
During training, the model parameters are optimized using a combination of reconstruction and adversarial loss. 
SoundStorm \cite{borsos2023soundstorm} is an improved version of SoundStream to achieve both efficiency and high-quality audio generation. 
It accomplishes this by employing an architecture specifically tailored to the hierarchical structure of audio tokens. 
Moreover, it pioneers a parallel, non-autoregressive decoding scheme, which relies on confidence-based strategies for residual vector-quantized token sequences. 

Encodec \cite{defossez2022high} builds upon a framework similar to SoundStream. 
Nonetheless, it further augments its capabilities by integrating supplementary LSTM \cite{hochreiter1997long} layers and harnessing a Transformer-based language model \cite{vaswani2017attention} to model the RVQ codes, thereby amplifying its sequence modeling performance.
Then, there is a stream of work aimed at making codec models more general and powerful.
AudioDec \cite{wu2023audiodec} represents an enhanced version of Encodec, implementing a group convolution mechanism to facilitate the real-time operation of the streamable network while also harnessing the capabilities of HiFi-GAN \cite{kong2020hifi} to effectively generate high-fidelity audio at a high sampling rate of 48 kHz.

In the AcademiCodec model introduced by \cite{yang2023hifi}, a novel technique known as group-residual vector quantization is presented. 
It employs multiple parallel RVQ groups.
This technique is specifically tailored for generation tasks. It aims to enhance the reconstruction performance while using a limited number of codebooks, consequently achieving an impressively low bit rate per second (BPS). 
This low BPS is of utmost significance as it effectively addresses the challenge of lengthy speech tokens in speech language modeling, resulting in reduced sequence lengths.

SpeechTokenizer \cite{zhang2023speechtokenizer} is a unified speech tokenizer designed for speech language models. 
It implements an Encoder-Decoder architecture enhanced with RVQ. 
By integrating both semantic and acoustic tokens, SpeechTokenizer hierarchically separates various aspects of speech information across different RVQ layers.
Specifically, SpeechTokenizer is designed to regularize the first RVQ layer to highlight semantic information by learning the Hubert tokens \cite{hsu2021hubert}.
Using such techniques can enhance the disentanglement of information across different RVQ layers.

Descript-audio-codec (DAC) \cite{kumar2023high}, a universal neural codec model, distinguishes itself through its exceptional ability to maintain high-fidelity audio quality across a wide spectrum of data types, encompassing general audio, music, and speech.
It accomplishes this feature by employing a number of training techniques, such as periodic activation functions \cite{ziyin2020neural}, enhanced residual vector quantization using factorized and L2-normalized codes, random quantizer dropout to preserve audio reconstruction quality, as well as refining adversarial and reconstruction loss during the training process.
The authors highlight the crucial importance of the periodic activation function among the employed techniques.

Unlike most models focusing on the time domain, FunCodec \cite{du2023funcodec} proposes a frequency-domain codec.
The authors claim they can achieve comparable performance with fewer parameters and lower computation complexity. 
Meanwhile, it also finds that incorporating semantic information in the codec tokens improves speech quality at low bit rates.

\subsection{Comparison from methodology angles}
% Evan: Describe the difference between discriminators. 
We compare several techniques proposed by these codecs in Table~\ref{tab:codec_compare}. The abbreviation ``A-F'' represents different codec models. 
Please refer to Table~\ref{tab:codec_info} for the corresponding model full name. 
The design of discriminators constitutes a pivotal element within codec models. 
Encodec initially introduces the Multi-scale-STFT Discriminator (MS-STFTD). 
In contrast to the multi-scale discriminator (MSD) proposed in MelGAN \cite{kumar2019melgan}, which captures long-term dependencies, the multi-period discriminator (MPD) proposed in HiFi-GAN \cite{kong2020hifigan} exhibits a capacity to discern more nuanced periodic details. 
Consequently, AudioDec replaces the conventionally employed STFTD with a HiFi-GAN-based MPD, observing an enhancement in audio quality within their model.
AcademiCodec integrates prior research efforts by incorporating the MS-STFTD from Encodec and both HiFi-GAN-based MPD and MSD. 
Both SpeechTokenizer and Funcodec adopt identical discriminators to AcademiCodec, with Funcodec offering a unified interface adaptable to any combination of these three discriminator types.
DAC identifies that employing MSD and MPD alone generates audio displaying blurriness and artifacts. 
To address this, they propose the application of a multi-scale, multi-band STFT discriminator (MS-MB-STFTD) to improve phase modeling and mitigate aliasing artifacts.

% Evan: Training techniques
SpeechTokenizer utilizes semantic tokens from Hubert L9 as a teacher for the RVQ process. 
This guidance enables the disentanglement of content information into the first layer of the tokenizer, while paralinguistic information is retained in subsequent layers.
FunCodec seeks to integrate semantic information by combining, adding, or residualizing the audio codec with semantic tokens. 
The study reveals that including semantic tokens enhances audio quality, particularly with the residual inclusion method. 
Additionally, SpeechTokenizer and FunCodec utilize K-means to cluster samples in the first mini-batch for initializing the VQ codebook, leading to improved code utilization.
% Furthermore, Funcodec is trained on Wenet for generalized purpose, a speech dataset with a higher degree of non-speech noise, resulting in a higher speech quality score on the Wenet test set compared to Encodec, DAC, and AudioDac. 
DAC follows the approach of BigVGAN \cite{lee2022bigvgan}, employing snake activation \cite{ziyin2020neural} for trainable control over the frequency of periodic signals. 
AcademiCodec employs multiple RVQ codebooks (multiple residual groups) to represent intermediate features. 
They demonstrate that using multiple residual groups achieves good reconstruction performance while employing only a few codebooks.
Encodec trains an additional small transformer model for entropy coding over the quantized units, which reduces bandwidth and accelerates encoding and decoding.

% Evan: Comparison of methodology
\begin{table}[t!]
\centering
\fontsize{8}{10}\selectfont
\setlength\tabcolsep{2pt}
\caption{Codec information comparison. "A-F" represents different neural codec models, where "A" is SpeechTokenizer \cite{zhang2023speechtokenizer}, "B$\sim$" is AcademiCodec \cite{yang2023hifi}, "C" is AudioDec \cite{wu2023audiodec}, "D$\sim$" is DAC \cite{kumar2023high}, "E$\sim$" is EnCodec \cite{defossez2022high}, and "F$\sim$" is FunCodec \cite{du2023funcodec}.  $n_c$ represents the codebook number, SR represents the sample rate, and BPS represents the bit rate in unit bits per second.}
\label{tab:codec_info}
\begin{tabularx}{0.48\textwidth}{lcccccc}
    \toprule
    & \textbf{Codec information} & \textbf{Training data} & $n_c$ & \textbf{SR} & \textbf{BPS} \\
    \midrule
    A & 16k & Librispeech & 8 & 16 & 4 \\
    \midrule
    B1 & hifi\_16k\_320d & LibriTTS & 4 & 16 & 2 \\
    B2 & hifi\_16k\_320d\_large\_uni & VCTK & 4 & 16 & 2 \\
    B3 & hifi\_24k\_320d & AISHELL & 4 & 24 & 3 \\
    \midrule
    C & 24k\_320d & Valentini  & 8 & 24 & 6.4 \\
    \midrule
    D1 & 16k & Common Voice, DAPS & 12 & 16 & 6 \\
    D2 & 24k & VCTK, MUSDB & 32 & 24 & 24 \\
    D3 & 44k & Jamendo, AudioSet & 9 & 44.1 & 8 \\
    \midrule
    E1 & 24k\_1.5bps  &   & 2 & 24 & 1.5 \\
    E2 & 24k\_3bps & Common Voice  & 4 & 24 & 3 \\
    E3 & 24k\_6bps & DAPS, Jamendo  & 8 & 24 & 6 \\
    E4 & 24k\_12bps & AudioSet, FSD50K  & 16 & 24 & 12 \\
    E5 & 24k\_24bps &  & 32 & 24 & 24 \\
    \midrule
    F1 & en\_libritts\_16k\_gr1nq32ds320 &  & 32 & 16 & 16 \\
    F2 & en\_libritts\_16k\_gr8nq32ds320  & Subset of LibriTTS & 32 & 16 & 16 \\
    F3 & en\_libritts\_16k\_nq32ds320 &  & 32 & 16 & 16 \\
    F4 & en\_libritts\_16k\_nq32ds640 &  & 32 & 16 & 8 \\
    % \cline{3-3}
    F5 & zh\_en\_16k\_nq32ds320 & 25k hours collected data & 32 & 16 & 16 \\
    F6 & zh\_en\_16k\_nq32ds640 & (en and zh-cn) & 32 & 16 & 8 \\
    \bottomrule
\end{tabularx}
\end{table}

\begin{table}[t!]
\centering
\fontsize{8}{10}\selectfont
\setlength\tabcolsep{2pt}
\caption{Comparison between codec implementation strategy. 
\textbf{SEM} represents codec including semantic tokens. \textbf{Snake} represents the codec model that employs snake activation. \textbf{MRG} represents codec has multiple residual groups. \textbf{noisy} represents codec utilizes noisy data in training. \textbf{LM} represents the model including language model training. \textbf{KM} represents codec uses K-means to cluster samples as initialization of VQ codebook.}
\label{tab:codec_compare}
\begin{tabularx}{0.48\textwidth}{lccccccc}
\toprule
 \textbf{Codec} & \textbf{Discriminators}       & \textbf{SEM} & \textbf{Snake} & \textbf{MRG} & \textbf{Noisy} & \textbf{LM} &\textbf{KM}\\
\midrule
A & MSD + MPD + MS-STFTD & \ding{51} & \ding{55} & \ding{55} & \ding{55} & \ding{55} & \ding{51} \\
B & MSD + MPD + MS-STFTD & \ding{55} & \ding{55} & \ding{51} & \ding{55} & \ding{55} & \ding{55} \\
C & MPD & \ding{55} & \ding{55} & \ding{55} & \ding{55} & \ding{55} & \ding{55} \\
D & MPD + MS-MB-STFTD & \ding{55} & \ding{51} & \ding{55}  &  \ding{55} & \ding{55} & \ding{55} \\
E & MS-STFTD & \ding{55} & \ding{55} & \ding{55} & \ding{55} &\ding{51} & \ding{55} \\
F & MSD + MPD + MS-STFTD & \ding{51} & \ding{55} & \ding{55} & \ding{51} & \ding{55} & \ding{51}   \\
\bottomrule
\end{tabularx}
\end{table}

% Evan: Comparison of implementation detail
\subsection{Implementation details}
We compare the codebook number, training data, sampling rate, and bit rate per second in Table~\ref{tab:codec_info}. 
% Due to computation resource limitation, we only choose the default BPS of these codecs in our experiment.
% Evan: Training Dataset brief intro.
From the training data perspective, SpeechTokenizer \cite{zhang2023speechtokenizer}, AudioDec \cite{wu2023audiodec}, and FunCodec \cite{du2023funcodec} utilize only English speech dataset. 
AcademiCodec \cite{yang2023hifi} incorporates bilingual speech datasets, including AISHELL for Chinese and LibriTTS and VCTK for English. 
Both DAC \cite{kumar2023high}, and Encodec \cite{defossez2022high} encompass diverse modality data, including speech, music, and audio, in the training data.

% Haibin: A figure with description: what is codec-based speech LM; refer to https://arxiv.org/pdf/2301.02111.pdf
\begin{figure}[t]
    \centering
    \includegraphics[width=0.8\columnwidth]{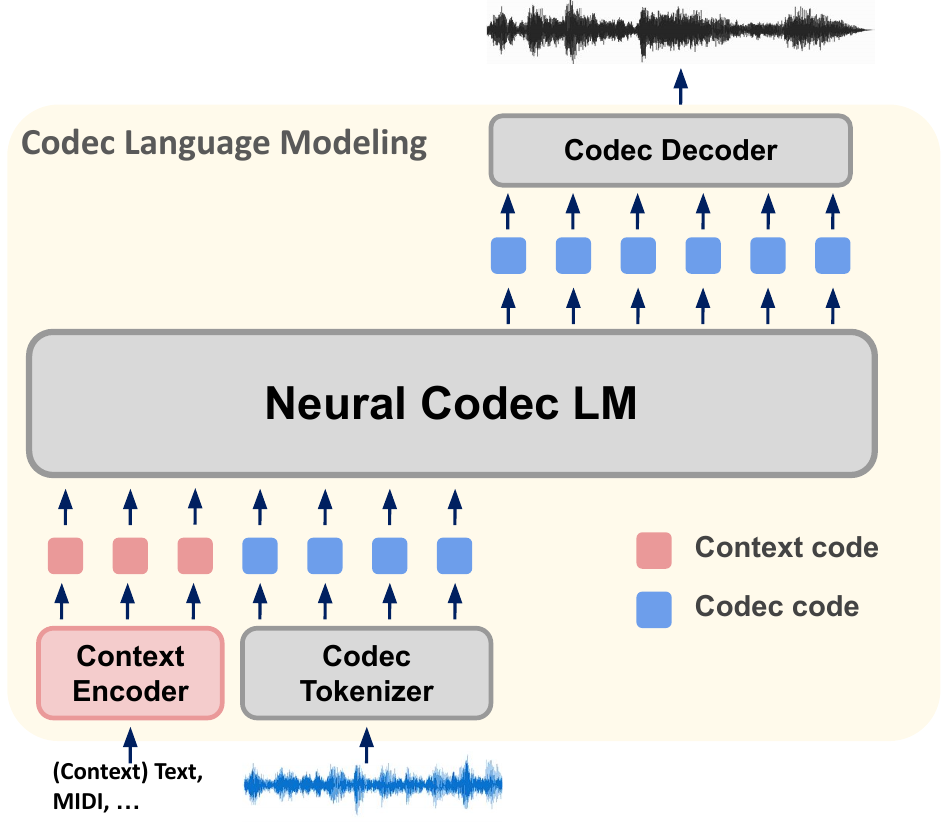}
    \caption{Codec-based Language Modeling}
    \vspace{-6pt}
    \label{fig:codec_lm}
\end{figure}

\section{Current codec-based speech language models}
% \subsection{Framework}

\noindent
As shown in Figure~\ref{fig:codec_lm}, the process of neural codec-based audio language modeling begins by converting context information, such as text and MIDI, into context codes, while simultaneously encoding the audio into codec codes. 
These context and codec codes are then employed in the language modeling phase to generate the desired target codec code sequence. 
Subsequently, the target codec code sequence is passed to the codec decoder to produce the audio output. The entire pipeline embodies an audio-to-audio modeling approach.

\subsection{Overview for codec-based LMs}
AudioLM \cite{borsos2023audiolm} is the pioneering model in introducing codec codes for language modeling, utilizing a hierarchical approach that encompasses two distinct stages.
The first stage generates semantic tokens using a self-supervised w2v-BERT model \cite{chung2021w2v}. 
These tokens are then leveraged in the second stage as conditioning elements to create acoustic tokens using a SoundStream neural codec \cite{zeghidour2021soundstream}.

VALL-E \cite{wang2023neural}, VALL-E X \cite{zhang2023speak}, and SpeechX \cite{wang2023speechx}, all originate from Microsoft and are neural codec language models trained to generate discrete codes derived from EnCodec \cite{defossez2022high}, based on textual or acoustic inputs.
VALL-E can generate high-quality personalized speech with only a 3-second enrollment recording from an unseen speaker. 
Furthermore, VALL-E X can produce high-quality speech in the target language with just a single speech utterance in the source language as a prompt. 
Additionally, SpeechX introduces a unified framework to address not only zero-shot TTS but also various types of speech transformation tasks, including speech enhancement and speech editing.

What sets ViaLA \cite{wang2023viola}, AudioPaLM \cite{rubenstein2023audiopalm}, and LauraGPT \cite{chen2023lauragpt} apart is their dual capability to generate both text and audio.
VioLA tries to tackle the question ``Is one decoder-only generative model all you need for speech recognition, synthesis, and translation?'' by employing language modeling that integrates both text tokens and audio tokens (extracted by EnCodec \cite{defossez2022high}), along with the use of task IDs and language IDs.
AudioPaLM constructs a unified vocabulary comprising both text and audio tokens. 
It is a decoder-only, autoregressive model capable of processing and generating both text and speech. 
Additionally, AudioPaLM's initialization stems from PaLM-2 \cite{anil2023palm}, a text-only language model.
AudioPaLM's approach to audio tokenization resembles that of AudioLM. 
Moreover, AudioPaLM adopts and extends the SoundStream model to SoundStorm \cite{borsos2023soundstorm}.
LauraGPT \cite{chen2023lauragpt} is a versatile language model built on a decoder-only text-based language model, Qwen-2B \cite{bai2023qwen}.
LauraGPT has the capability to process both audio and text inputs, generating outputs in either modality. 
LauraGPT encodes input audio into continuous representations using a Conformer encoder and decodes output audio using FunCodec \cite{du2023funcodec} discrete codes. 
The authors claim this specific audio features design for inputs and outputs will result in improved performance for speech generation using some preliminary experimental results.

UniAudio \cite{yang2023uniaudio} utilizes language modeling to generate a wide range of audio types, including speech, sounds, music, and singing, using textual or acoustic tokens as inputs. 
UniAudio stands out for its ability to enhance autoregressive prediction speed by introducing a multi-scale Transformer model \cite{yu2023megabyte}, which employs a large global transformer to predict the first-layer codec codes and a small local transformer to predict the codec codes for the subsequent codec layers.
The codec model in UniAudio is revised from EnCodec.

Additionally, there are other codec-based language models designed for sound modeling.
AudioGen \cite{kreuk2022audiogen} trained a SoundStream model to get audio tokens and subsequently trained a language model to utilize textual features as conditions for generating audio tokens.
MusicLM \cite{agostinelli2023musiclm} follows a training strategy similar to AudioLM but extends its scope to encompass music features.
It approaches the task of conditional music generation through a hierarchical sequence-to-sequence modeling approach. 
Initially, it utilizes music tokens from Mulan \cite{huang2022mulan} to generate semantic tokens from the w2v-BERT model. 
Subsequently, it employs both music tokens and semantic tokens to generate acoustic features through Soundstream.
MusicGen \cite{copet2023simple} is a music language model designed to work with EnCodec discrete tokens. 
It accepts textual descriptions or melodic features as input conditions to generate tokens, which can be reconstructed to high-fidelity music.

% Xuanjun
\begin{table}[t!]
\centering
\fontsize{7.3}{10}\selectfont
\setlength\tabcolsep{2pt}
\caption{Codec-based language models comparison. "T" means text, "AUD" means audio, "P" means phoneme, and "M" means MIDI.}
\label{tab:clm_model}
\begin{tabularx}{0.48\textwidth}{lcccc}
\toprule
\multirow{2}{*}{\textbf{CLM}} &  \multirow{2}{*}{\textbf{Task}} & \multirow{2}{*}{\textbf{Input}} & \multirow{2}{*}{\textbf{Output}} & \multirow{2}{*}{\textbf{Codec}}  \\
 & & & & ~ \\
\midrule
% AudioLM & Speech continuation & Semantic, acoustic & \multirow{11}*{Audio} \\
% AudioLM & SC, PC & Semantic, acoustic & \multirow{14}{*}{Audio}\\ 
AudioLM \cite{borsos2023audiolm} & SC, PC & AUD & AUD & SoundStream \cite{zeghidour2021soundstream} \\ 

AudioGen \cite{kreuk2022audiogen} & AC & AUD, T &  AUD & SoundStream \cite{zeghidour2021soundstream}  \\

VALL-E \cite{wang2023neural} & TTS & AUD, T & AUD & EnCodec \cite{defossez2022high} \\

MusicLM \cite{agostinelli2023musiclm} & MG & AUD, T &  AUD & SoundStream \cite{zeghidour2021soundstream}  \\

VALL-E X \cite{zhang2023speak} & TTS, S2ST & AUD,T & AUD &  EnCodec \cite{defossez2022high} \\

VioLA \cite{wang2023viola} & ASR, S2TT, TTS, MT & AUD,T &  AUD,T & EnCodec \cite{defossez2022high} \\ 
MusicGen \cite{copet2023simple}  & MG, SG & AUD & AUD & EnCodec \cite{defossez2022high} \\

AudioPaLM \cite{rubenstein2023audiopalm} & ASR, S2TT, TTS, MT & AUD,T & AUD,T & SoundStorm \cite{borsos2023soundstorm} \\ 

SpeechX \cite{wang2023speechx} & \makecell{SE, SR, TSE,  TTS, SPED} & AUD,T & AUD & EnCodec \cite{defossez2022high} \\

\midrule
\multirow{2}{*}{LauraGPT \cite{chen2023lauragpt}} & \multirow{2}{*}{\shortstack{ASR, S2TT, TTS, MT, SE\\ AAC, SER, SLU
}} & \multirow{2}{*}{AUD,T} &  \multirow{2}{*}{AUD, T} &  \multirow{2}{*}{FunCodec \cite{du2023funcodec}} \\
 & & & ~ \\
 
\midrule

\multirow{3}{*}{UniAudio \cite{yang2023uniaudio}} &  \multirow{3}{*}{\shortstack{TTS, VC, SE, TSE, SVS \\ TTSO, TTM, AUED \\ SD, ITTS, SPED}} & \multirow{3}{*}{\shortstack{P, M, \\ AUD,T}} &  \multirow{3}{*}{AUD} & \multirow{3}{*}{EnCodec \cite{defossez2022high}} \\ 
 & & & ~ \\
 & & & ~ \\

% \multirow{3}{*}{\shortstack{TTS, VC, SE, TSE, SVS \\ TTSO, TTM, AUED \\ SD, ITTS, SPED}}
\bottomrule
\end{tabularx}

\end{table}

Another branch of speech language modeling aims to utilize discrete units obtained by quantizing self-supervised speech representations.
While these discrete units contain rich acoustic and linguistic information~\cite{wells2022phonetic}, they lack speaker and paralinguistic information~\cite{DBLP:conf/interspeech/PolyakACKLHMD21}.
This research direction focuses on modeling the semantics of speech, with the optional use of encoders to learn about speaker characteristics and prosody.
Pioneering work is speech-resynthesis~\cite{DBLP:conf/interspeech/PolyakACKLHMD21}, which utilizes these discrete units in conjunction with prosody and speaker encoders to encode speech into low-bitrate codes. 
These codes can then be resynthesized into a speech signal with a decoder to achieve low-bitrate transmission.
Additionally, these discrete units can be regarded as ``pseudo-text,” serving as a foundation for training textless speech language models. 
Notable examples include GSLM~\cite{lakhotia2021generative}, pGSLM~\cite{kharitonov2021text}, dGSLM~\cite{nguyen2023generative}, and TWIST~\cite{hassid2023textually}. 
By engaging in the pre-trained task of next-token prediction, these speech LMs perform spoken language modeling and can conduct the task of speech continuation.
In the field of speech translation, recent advancements have been made possible through these discrete units. 
\cite{popuri22_interspeech} pre-trained a Unit mBART combined with a wav2vec 2.0 ~\cite{baevski2020wav2vec} encoder to directly predict the translated discrete units. 
UnitY~\cite{inaguma2022unity} further incorporates text modality to enhance speech translation. 
The Seamless models~\cite{barrault2023seamlessm4t, barrault2023seamless} integrate the UnitY framework to perform expressive and streaming speech-to-text and speech-to-speech translation.
With the development of these powerful speech LMs, researchers have begun to explore the use of prompting on speech LMs for various speech processing tasks, including prompt tuning~\cite{chang22e_interspeech, chang2023speechprompt,wu2023speechgen}, in-context learning~\cite{hsu2023exploration}, and instruction tuning~\cite{kuan2023towards, huang2023dynamic}.

\subsection{Comparison for Codec-based audio language models}

% Xuanjun
In Table~\ref{tab:clm_model}, we compare the inputs, outputs, and downstream tasks of different codec-based language models. 
We also summarize that the downstream tasks conducted by different codec-based language models: Speech Continuation (SC), Piano Continuation (PC), Audio Continuation (AC), Text-to-Speech (TTS), Music Generation (MG), Stereophonic Generation (SG), Speech to Speech Translation (S2ST), Automatic Speech Recognition (ASR), Spoken Language Understanding (SLU), Automated Audio Captioning (AAC), Speech to Text Translation (S2TT), Machine Translation (MT), Speech Enhancement (SE), Speech Removal (SR), Target Speaker Extraction (TSE), Speech Editing (SPED), Voice Conversion (VC), Singing Voice Synthesis (SVS), Text-to-Sound (TTSO), Text-to-Music (TTM), Audio Editing (AUED), Speech Dereverb (SD), Instructed TTS (ITTS).
Finally, we show the codec models adopted by different LMs.

\section{Conclusion}

The paper fills the research blank to review the neural codec models and LMs built upon them. 
We hope the comprehensive review and comparisons can inspire future research works to boost the development of neural codec models and codec-based LMs. 
% Our future work is to better integrate the advantages of different existing codecs to train better codec models and codec-based language models.

% \newpage
% \subsection{Information for Authors}
% \section{References}
\bibliographystyle{IEEEbib}

\bibliography{main}
\vfill\pagebreak

\end{document}